\documentclass[doublecol]{epl2}
\begin{document}

\title{Detrended cross-correlations between returns, volatility, trading activity, and volume traded for the stock market companies}

\author{Rafa{\l} Rak\inst{1}, Stanis{\l}aw Dro\.zd\.z\inst{2,3}, Jaros{\l}aw Kwapie\'n\inst{2}, \and Pawe{\l} O\'swi\c ecimka\inst{2}}
\shortauthor{R.~Rak \etal}

\institute{
\inst{1} Faculty of Mathematics and Natural Sciences, University of Rzesz\'ow, Rzesz\'ow, Poland \\
\inst{2} Institute of Nuclear Physics, Polish Academy of Sciences, ul. Radzikowskiego 152, Krak\'ow, Poland \\
\inst{3} Faculty of Physics, Mathematics and Computer Science, Cracow University of Technology, Krak\'ow, Poland}


\abstract{
We consider a few quantities that characterize trading on a stock market in a fixed time interval: logarithmic returns, volatility, trading activity (i.e., the number of transactions), and volume traded. We search for the power-law cross-correlations among these quantities aggregated over different time units from 1 min to 10 min. Our study is based on empirical data from the American stock market consisting of tick-by-tick recordings of 31 stocks listed in Dow Jones Industrial Average during the years 2008-2011. Since all the considered quantities except the returns show strong daily patterns related to the variable trading activity in different parts of a day, which are the best evident in the autocorrelation function, we remove these patterns by detrending before we proceed further with our study. We apply the multifractal detrended cross-correlation analysis with sign preserving (MFCCA) and show that the strongest power-law cross-correlations exist between trading activity and volume traded, while the weakest ones exist (or even do not exist) between the returns and the remaining quantities. We also show that the strongest cross-correlations are carried by those parts of the signals that are characterized by large and medium variance. Our observation that the most convincing power-law cross-correlations occur between trading activity and volume traded reveals the existence of strong fractal-like coupling between these quantities.}

\maketitle

\section{Introduction}

Complex systems are typically characterized by a hierarchical structure, in which the hierarchy is observed both as different levels of the system's organization from a micro scale to progressively higher macro scales (for example, from molecules to the multicell organisms and ecosystems) and as the structures of different size that coexist at the same organization level (for example, organisms with different body size or species with different population). This same-level hierarchy is often observed indirectly as the power-law distributions or the power-law temporal correlations of measurement outcomes. The latter are especially interesting from the physical point of view since such behaviour (i.e., long memory) can be related to critical phenomena in the system's evolution~\cite{kwapien2012}. Financial markets are perhaps the most complex structures in the world, which makes the related data to be a particularly interesting object of studying. Here we focus on the stock market data and the most fundamental quantities: price, trades, and volume traded. 

Different studies showed that there is a relation between traded volume $V$ and a price change $\Delta p(t)= p(t+\Delta t)-p(t)$ (or a logarithmic return $r(t) = \ln (p(t+\Delta t)/p(t)$) of this stock over the same time interval $\Delta t$. This relation (which should not necessarilly be causal) concerns both the signed price change $\Delta p(t)$ or the return $r(t)$ and the absolute price change (volatility): $|\Delta p(t)|$, $|r(t)|$~\cite{ying1966,crouch1970}. An exact form of this relation is still a subject of debate since a number of studies devoted to this issue provided us with mixed results, but its very existence seems to be well established now. As regards its origin, there is no decisive conclusion yet. It might well be a consequence of a relation between volume and trading activity (the number of transactions in a unit time interval), between volume and transaction size, or an effect of the fluctuations in order imbalance~\cite{jones1994,chan2000,lillo2003,potters2003,bouchaud2004,lim2005,zhou2012,rak2013}.

A specific type of coupling between different quantities is power-law cross-correlations. They may occur between the quantities that are power-law autocorrelated, which is the case in the financial market data~\cite{kwapien2012}. The cross-correlations between absolute fluctuations of the S\&P500 index and the corresponding volume traded were discovered in~\cite{podobnik2009b}. No such correlations were found between the volume and price fluctuations, however. In contrast, an analysis of daily price and volume fluctuations of the indices of the Chinese main stock markets located in Shanghai (SHSI) and Shenzhen (SZCI) showed that these quantities do depend on each other and that this dependence is power-law and multifractal~\cite{yuan2012}. Similar outcomes were obtained from an analysis of the high-frequency price and volume fluctuations of the Chinese CSI300 index futures with an additional indication that these cross-correlations were persistent~\cite{wang2013}.

Here we study cross-correlations among the following quantities: price returns, volatility, trading activity, and volume traded. We analyze the tick-by-tick data representing highly capitalized stocks that were listed in the Dow Jones Industrial Average (DJIA) during the period 2008-2011. What distinguishes our study is that we remove daily trends from all types of data before we proceed with the analysis. We investigate the power-law cross-correlations by means of the multifractal detrended cross-correlation analysis with sign preserving (MFCCA~\cite{oswiecimka2014}) that is a consistent generalization of the detrended cross-correlation approach (DCCA~\cite{podobnik2008}). According to our knowledge, this is the first analysis of this type that incorporates such a set of quantities and approach.

\section{Data}

Let us assume that one records every trade in which $\omega^{(u)}(t_i)$ shares of a stock $u$ ($u = 1,...,N$) at price $p^{(u)}(t_i)$ are transfered between a pair of investors, where $t_i$ is the moment of the $i$th trade. Let us also assume that consecutive time intervals of length $\Delta t$ are indexed by $j$ ($j = 1,...,j_{\rm max}$) and that the effects of all the trades $t_i$ ($i=1,..,T^{(u)}(j)$) that were made during each such an interval are aggregated. Thus, for each interval $j$ we may consider the following four quantities: (i) the logarithmic price returns:
\begin{equation}
R^{(u)}(j) = \ln p^{(u)}(j \Delta t) - \ln p^{(u)}((j-1) \Delta t)
\end{equation}
(where we assumed that the first interval begins at $t=0$ and that the price remains unchanged between the trades), (ii) the trading activity $T^{(u)}(j)$, and (iii) the total volume traded
\begin{equation}
V^{(u)}(j) = \sum_{i=1}^{T^{(u)}(j)} \omega^{(u)}(t_i).
\end{equation}
In addition, we define volatility of an interval $j$ as $|R^{(u)}(j)|$.

The dataset under our study consists of the tick-by-tick recordings representing $N=31$ stocks that were listed (at least temporarily) in DJIA over the period from Jan 1, 2008 to Jul 31, 2011. For each stock, we select three time lags $\Delta t$ equal to 1, 5, and 10 min and for each lag we create four parallel time series of length $L \approx 3 \times 10^5$, $L \approx 6 \times 10^4$, and $L \approx 3 \times 10^4$, respectively, corresponding to the variables listed above. These variables, especially the returns and volume traded, are known to exhibit power-law tails of the probability distribution functions for short time lags $\Delta t$ of order of seconds and minutes that tend to depart from power law for longer $\Delta t$, what was documented earlier in~\cite{drozdz2003,drozdz2007,rak2007} and recently reconfirmed in~\cite{botta2015} for similar data set as in our present work. Next, we remove daily trends from each of the unsigned time series $T^{(u)}$, $V^{(u)}$, $|R|^{(u)}$ by a standard procedure~\cite{liu1999}, in which the daily pattern $D_Z^{(u)}$ is calculated as an arithmetic average of the respective quantity $Z$ taken at a specific interval $j_d$ of each trading day $k$ over all $N_{\rm days}$ trading days:
\begin{equation}
D_Z^{(u)}(j_d) = {\sum_{k=1}^{N_{\rm days}} Z^{(u)}(k,j_d) \over N_{\rm days}}.
\end{equation}
It is removed by division:
\begin{equation}
Z_{\rm detr.}^{(u)}(k,j_d) = Z^{(u)}(k,j_d) / D_Z^{(u)}(j_d).
\end{equation}
For simplicity of the notation, we will omit the superscripts $(u)$ in $R^{(u)}$, $|R^{(u)}|$, $T^{(u)}$, and $V^{(u)}$ henceforth.

\section{Power-law detrended autocorrelations}

Such autocorrelations can be detected by applying the multifractal detrended correlation analysis that is defined as follows~\cite{kantelhardt2002}. Let there be a time series $x(i)_{i=1,...,L}$ divided into $2 M_s$ separate boxes of length $s$ (i.e., $M_s$ boxes starting from the opposite ends). In each box $\nu$ ($\nu = 0,...,2 M_s - 1$) a local trend of an integrated signal is approximated by an $m$th-degree polynomial $P^{(m)}$ and removed by subtraction:
\begin{equation}
X_{\nu}(s,i) = \sum_{j=1}^i x(\nu s +j) - P^{(m)}_{X,s,\nu}(j)
\label{eq::detrended.signal}
\end{equation}
producing a detrended signal $X_{\nu}$ of length $s$ (we will use $m=2$ throughout this study). Variance of this signal can be expressed by
\begin{equation}
f^2(s,\nu) = {1 \over s} \sum_{i=1}^s X_{\nu}^2(s,i)
\label{eq::variance}
\end{equation}
and can be used to define a family of the fluctuation functions of order $q$:
\begin{equation}
F^q(s) = {1 \over 2 M_s} \sum_{\nu=0}^{2 M_s - 1} \left[ f^2(s,\nu) \right]^{q/2}.
\label{eq::fluctuation.function.mfdfa}
\end{equation}
The fluctuation functions can be used to detect fractality of the signal $x(i)$ if they obey a power-law form: $F^q(s) \sim s^{\gamma(q)}$, where $\gamma(q)$ can either be linear or nonlinear. In the former case, $x(i)$ is called monofractal, while in the latter case it is called multifractal. If one prefers to look at a graphical representation of the fluctuation functions, it is more instructive to use their modified versions:
\begin{equation}
F_q(s) = \left[ F^q(s) \right]^{1/q}
\end{equation}
since in that case one can easily distinguish between a monofractal and a multifractal scaling: $F_q(s) \sim s^{h(q)}$,  where $h(q)$ is constant for a monofractal signal and variable for a multifractal one.

\begin{figure}
\includegraphics[scale=0.32]{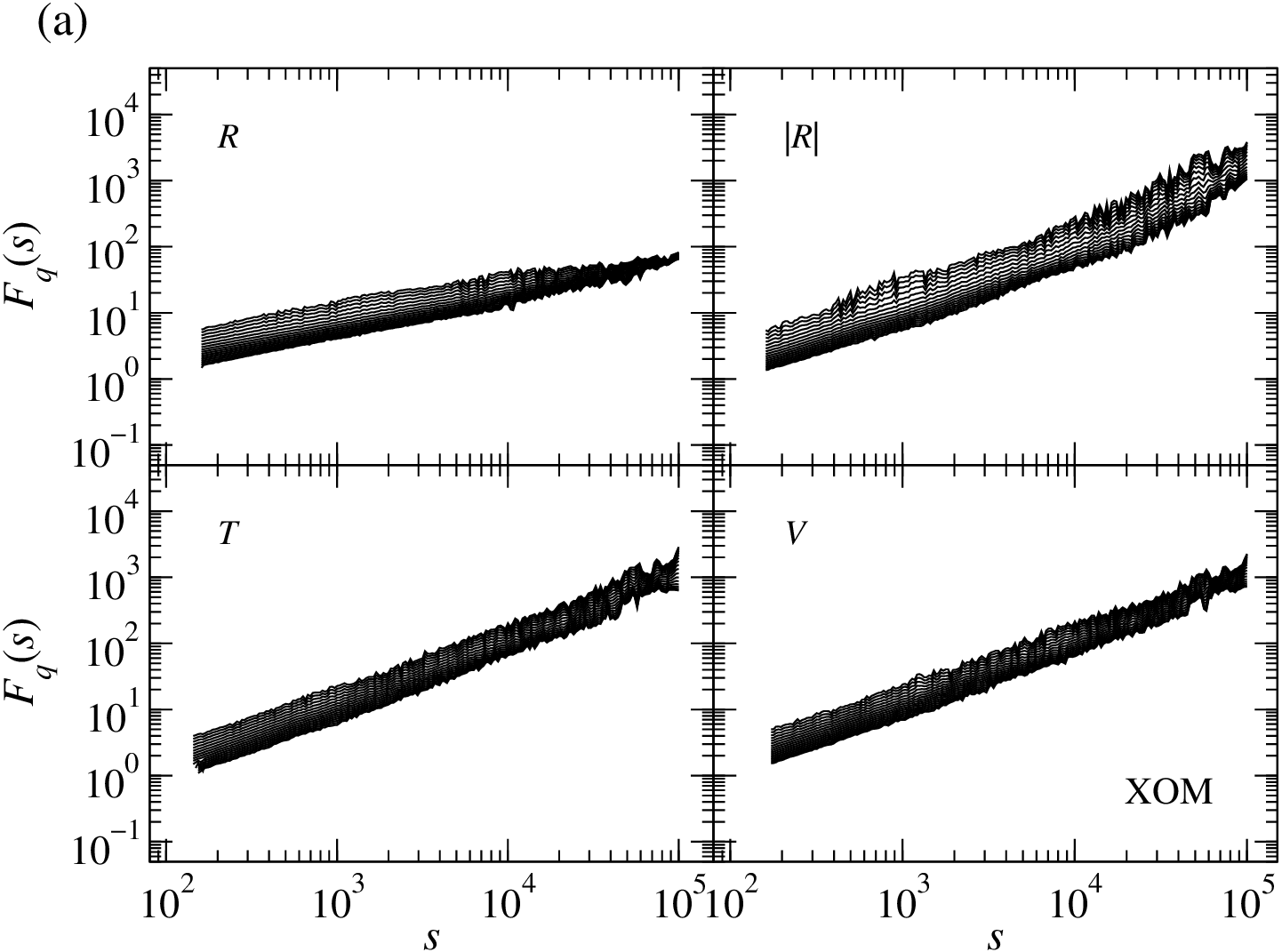}

\vspace{0.8cm}
\includegraphics[scale=0.32]{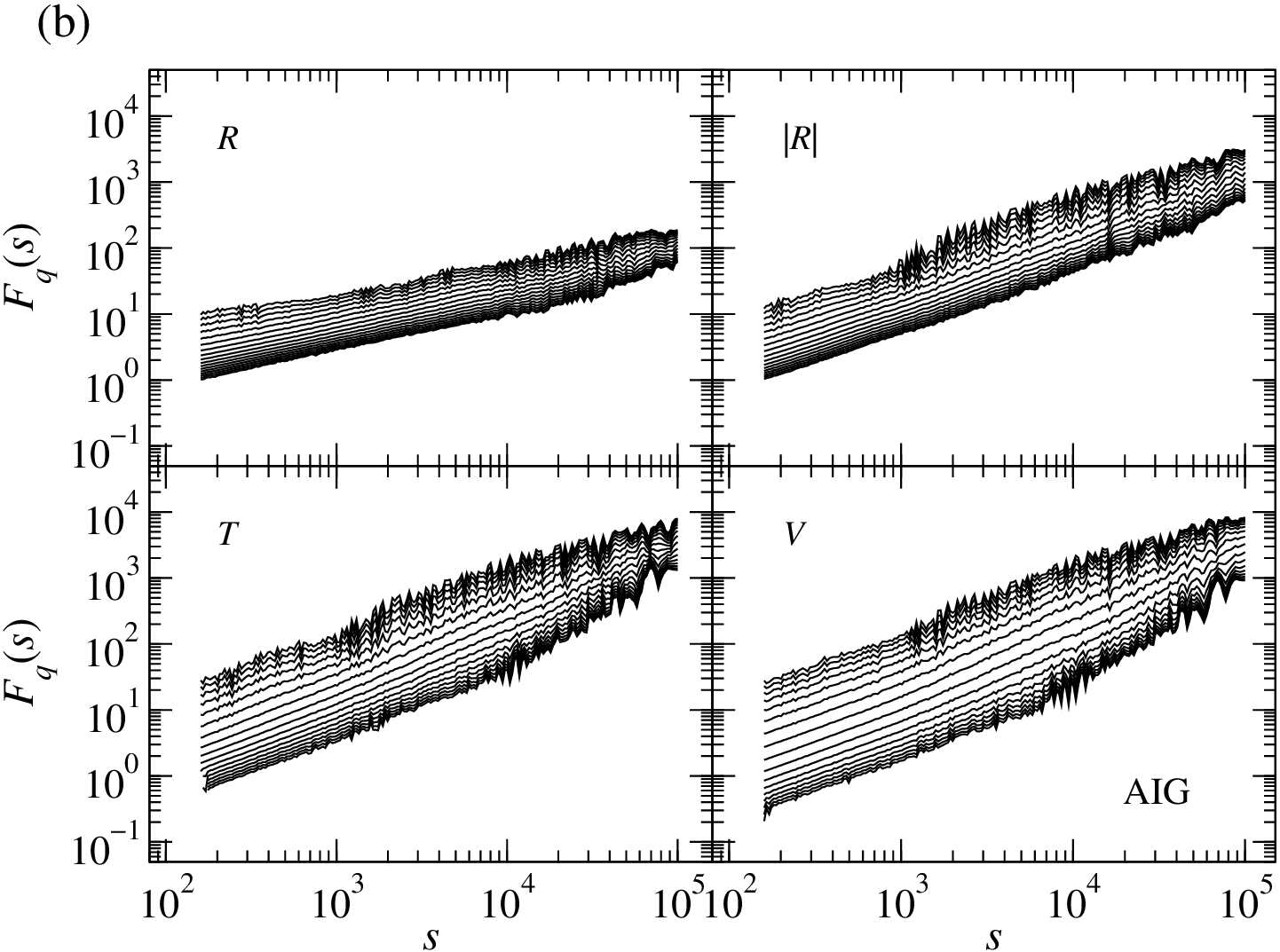}
\caption{The fluctuation functions $F_q(s)$ for two sample stocks with different scaling properties: XOM (a) and AIG (b). Results obtained for different types of data are shown in different panels: (top left) returns $R$, (top right) volatility $|R|$, (bottom left) trading activity $T$, and (bottom right) volume traded $V$. In each case the parameter $q$ is restricted to the interval $\langle -4,4 \rangle$ with a step of 0.5 (lines from bottom to top).}
\label{fig::mfdfa.fractal}
\end{figure}

Fig.~\ref{fig::mfdfa.fractal} shows the fluctuation functions $F_q(s)$ calculated for sample stocks with different types of scaling: XOM and AIG. One can see that all types of signals reveal at least approximate scaling but sometimes for different scale ranges. We obtain the same result for other considered stocks as well. The multifractal scaling is a well-known feature of the stock-market data~\cite{ivanova1999,dimatteo2003,oswiecimka2005,muzy2008,calvet2002,drozdz2009,drozdz2010,ludescher2011a}, therefore we consider possible fractal cross-correlations among these signals, instead.

\section{Power-law detrended cross-correlations}

We proceed along the line defined by the multifractal detrended cross-correlation analysis in its MFCCA variant~\cite{oswiecimka2014}. It is distinct from MFDFA in that that here one deals with two detrended signals: $X_{\nu}(s,i)$ and $Y_{\nu}(s,i)$ in each box $\nu$ instead of only one signal in Eq.~(\ref{eq::detrended.signal}). Variance defined by Eq.~(\ref{eq::variance}) becomes covariance:
\begin{equation}
f^2_{XY}(s,\nu) = {1 \over s} \sum_{i=1}^s X_{\nu}(s,i) Y_{\nu}(s,i)
\label{eq::covariance}
\end{equation}
and can have negative values now. The respective $q$th-order fluctuation function is then expressed by the following formulae:
\begin{equation}
F_{XY}^q(s) = {1 \over 2 M_s} \sum_{\nu=0}^{2 M_s - 1} {\rm sign} \left[ f_{XY}^2 (s,\nu) \right] |f_{XY}^2(s,\nu)|^{q/2}
\label{eq::fluctuation.function.mfcca}
\end{equation}
and
\begin{equation}
F_{qXY}(s) = {\rm sign} \left[ F_{XY}^q(s) \right] |F_{XY}^q(s)|^{1/q}.
\label{eq::fluctuation.function.mfcca.modified}
\end{equation}
Definition of the functions in Eqs.~(\ref{eq::fluctuation.function.mfcca}) and (\ref{eq::fluctuation.function.mfcca.modified}) assures us that no imaginary parts may be obtained by raising $f^2_{XY}$ to a real power $q/2$ and $F_{XY}^q$ to a real power $1/q$, and also assures us that no information about the signs is lost by taking the moduli. Moreover, for $q=2$ the MFCCA procedure consistently transforms itself into the basic detrended cross-correlation analysis (DCCA)~\cite{podobnik2008}, exactly as desired. A fractal character of the cross-correlations is related to a power-law scaling: $F_{XY}^q(s) \sim s^{\delta(q)}$ and $F_{qXY}(s) \sim s^{\lambda(q)}$, where $\delta(q) \sim q$ and $\lambda(q) = const$ denote monofractality. (We stress the necessity of defining $F_{XY}^q$ exactly as in Eq.~(\ref{eq::fluctuation.function.mfcca}), because the frequent, published attempts to avoid complex numbers by taking moduli of the detrended signals in Eq.~(\ref{eq::covariance}) lead to spurious multifractal scaling for virtually all the power-law autocorrelated signals even if they are mutually uncorrelated by construction~\cite{oswiecimka2014}.)

\begin{figure}
\includegraphics[scale=0.32]{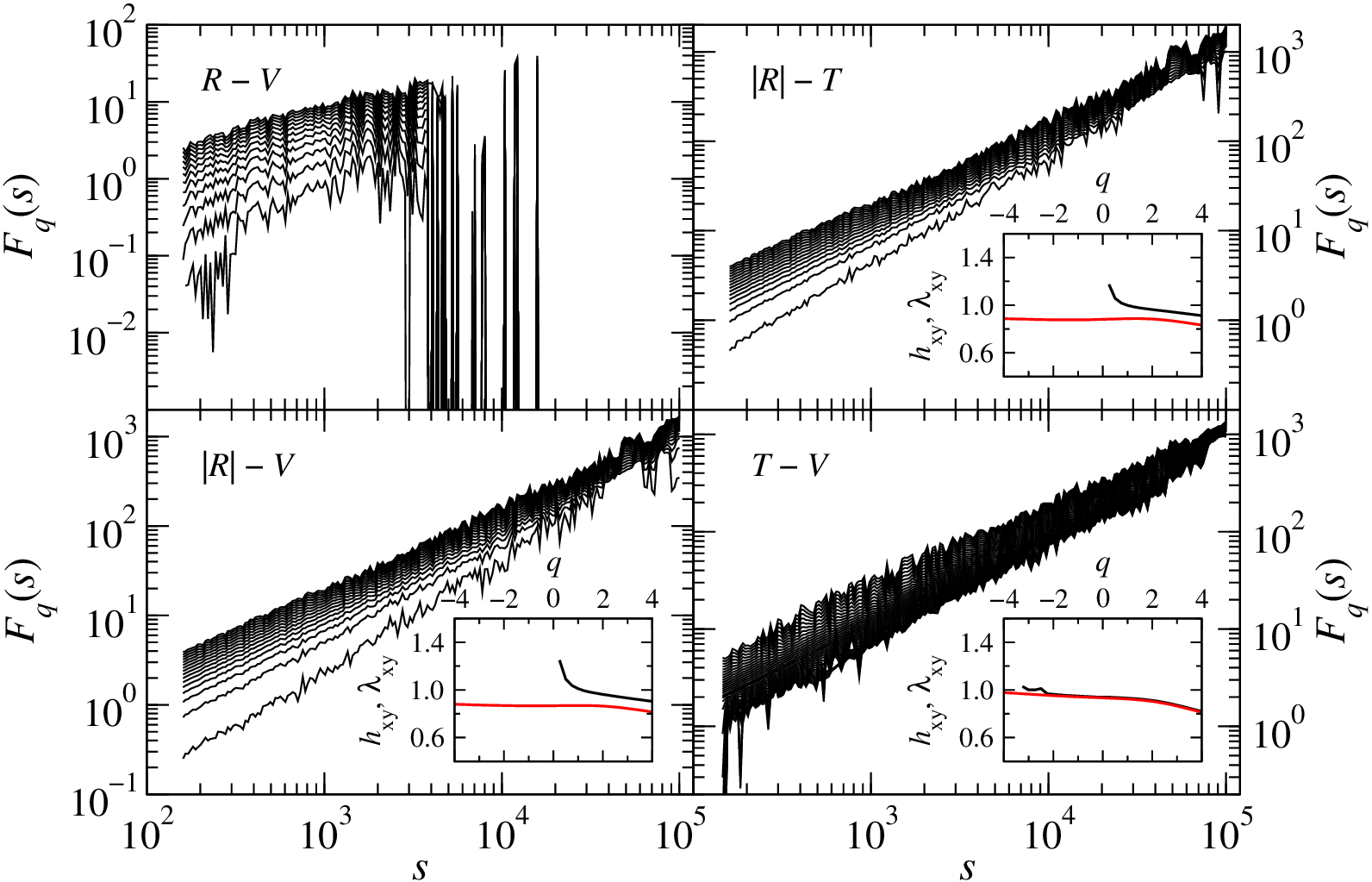}
\caption{The cross-correlation fluctuation functions $F_{qXY}(s)$ (main plots) and the scaling exponents: $h_{xy}(q)$ (red/grey, insets) and $\lambda(q)$ (black, insets) calculated for different pairs of 1-min time series representing the stock WMT: (a) returns and volume traded, (b) volatility and trading activity, (c) volatility and volume, and (d) trading activity and volume. Only the results for those values of $q$ are shown, for which $F_{qXY}(s)$ exhibits at least approximate scaling. The same range of $q$ is used as in Fig.~\ref{fig::mfdfa.fractal} with $q=4$ denoted by the topmost lines.}
\label{fig::mfcca.different.types}
\end{figure}

For each stock, we apply the MFCCA procedure to signal pairs representing 6 combinations of the 4 data types. The results show that there are significant differences in the character of the power-law cross-correlations between different pairs of signals and that these differences are consistent across all the stocks. In Fig.~\ref{fig::mfcca.different.types} we present plots of the fluctuation functions $F_{qXY}(s)$ calculated for different pairs of signals representing the stock WMT. This stock is characterized by the power-law autocorrelations existing in each of the 4 signals: $R$, $|R|$, $T$, and $V$, so it is well suited for serving as a representative example. Out of the studied data types, the returns are the only ones that are signed, so it is not surprising that the returns are the least cross-correlated with the other signals. Fig.~\ref{fig::mfcca.different.types}(top left) displays $F_{qXY}(s)$ for the pair $R$-$V$, while the remaining two pairs: $R$-$|R|$ and $R$-$T$ look qualitatively similar: the power-law behaviour is poor (if any) and restricted to $q > 0$ and to approximately one decade of scales (not shown here). Consequently, the scaling exponent $\lambda(q)$ cannot be defined despite the average generalized Hurst exponent:
\begin{equation}
h_{xy}(q) = {h_x(q) + h_y(q) \over 2},
\end{equation}
is defined for the whole analyzed range of $q$.

The next two pairs: $|R|$-$T$ and $|R|$-$V$ are similar to each other (Fig.~\ref{fig::mfcca.different.types} (top right) and (bottom left)). In both cases, $F_{qXY}(s)$ is a well-defined, two-and-a-half-decade wide power-law for $q>0$, while it is wildly unstable for $q \le 0$ and fluctuates from positive to negative values. This means that the signals are fractal cross-correlated indeed, but these correlations are restricted mainly to large fluctuations of $|R|$, $T$, and $V$. This result goes in parallel with the results of an earlier study~\cite{podobnik2009b} that also proved the existence of such cross-correlations in a market index. Both the variable slope of the fluctuation functions and the decreasing values of the scaling exponent $\lambda(q)$ indicate that the cross-correlations might be multifractal but this multifractality is rather subtle (a small range of the $\lambda(q)$ variability).

The cross-correlations that comprise both the small and the large fluctuations can be seen in $T$-$V$ (Fig.~\ref{fig::mfcca.different.types}(bottom right)). Here $F_{qXY}(s)$ are power-law correlated for all the considered values of $q$ and for more than two decades of the scales. This means that $T$-$V$ is the pair that retains the fractal structure of their autocorrelations in their cross-correlations the best. Consistently, it is also the only pair, in which $h_{xy}(q)$ is followed by $\lambda(q)$ (inset to Fig.~\ref{fig::mfcca.different.types}(bottom right)). Such strong power-law correlations in the properties of trading activity and volume support assumptions of the Gabaix et al.'s model of the mechanism that produces power-law distributions in the financial data~\cite{gabaix2003,gabaix2006}.

For trading activity and volume traded, the variability range of the scaling exponents $\lambda(q)$ for $q \in \langle-4,4\rangle$ is about 0.2, which suggests weak multifractality of their cross-correlations. On the other hand, even larger values of $\Delta \lambda(q)$ for the $|R|$-$T$ and $|R|$-$V$ pairs originate mainly from the strong variability of $\lambda(q)$ for small positive values of $0 < q < 1$, but this range seems too narrow to allow us to consider such a result to be a proof of true multiscaling.

\begin{figure}
\includegraphics[scale=0.32]{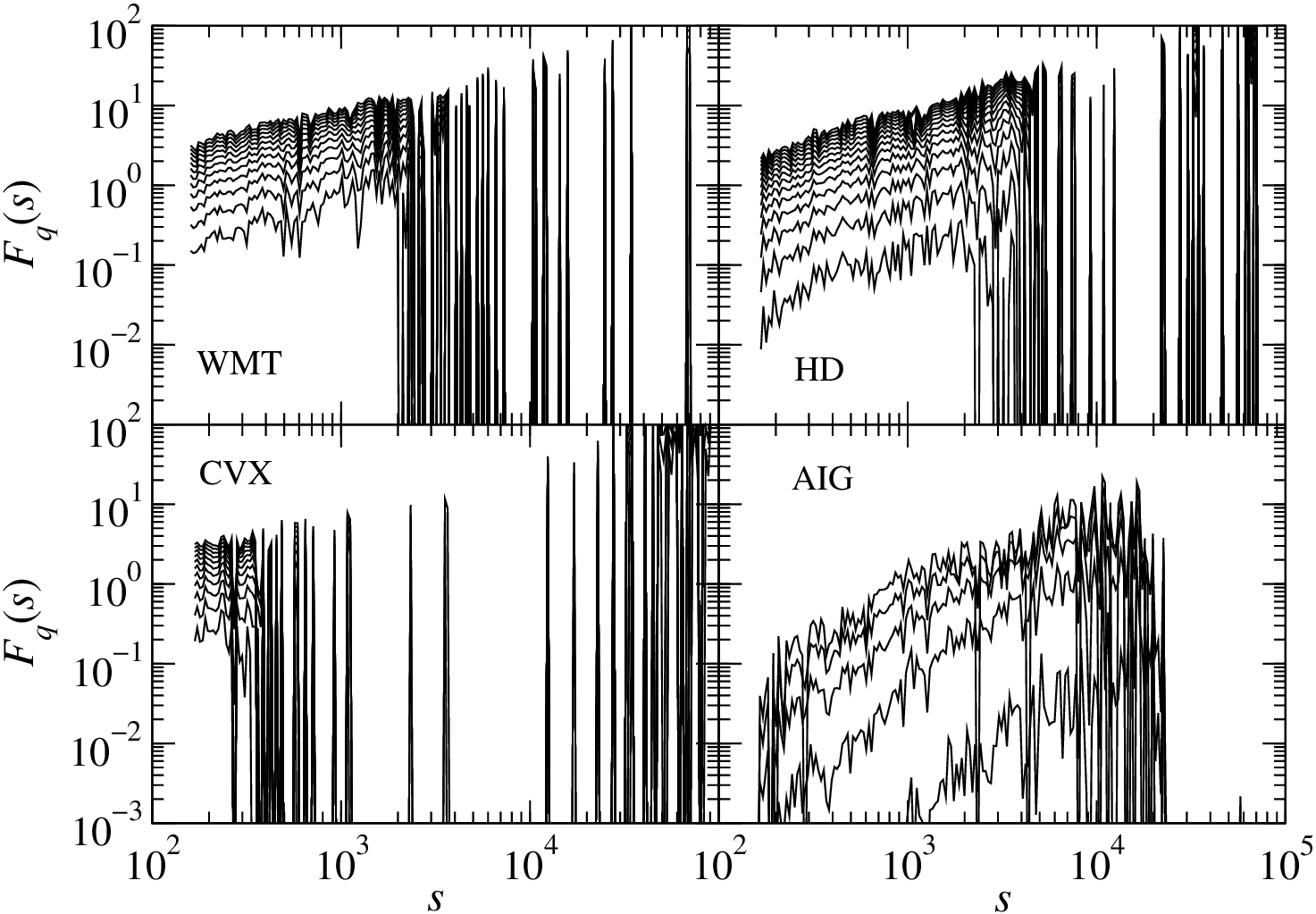}
\caption{The cross-correlation fluctuation functions $F_{qXY}(s)$ for the pair $R$-$V$ with $\Delta t=1$ min. Each panel corresponds to a different stock. In each case the topmost line corresponds to $q=4$ and the subsequent lines correspond to smaller values of $q$ with a step of $\Delta q=-0.5$.}
\label{fig::mfcca.rv}
\end{figure}

Fig.~\ref{fig::mfcca.rv} shows representative results for the pair $R$-$V$ for a few stocks. They vary from the poor, narrow-range power-law-like dependence in $F_{qXY}(s)$ for WMT and HD to the unstable, non-scaling functional dependences for CVX and AIG. It is worth recalling here that this lack of the power-law cross-correlations is accompanied by the well-established power-law autocorrelations in both $R$ and $V$ (compare Fig.~\ref{fig::mfdfa.fractal}).

\begin{figure}
\includegraphics[scale=0.32]{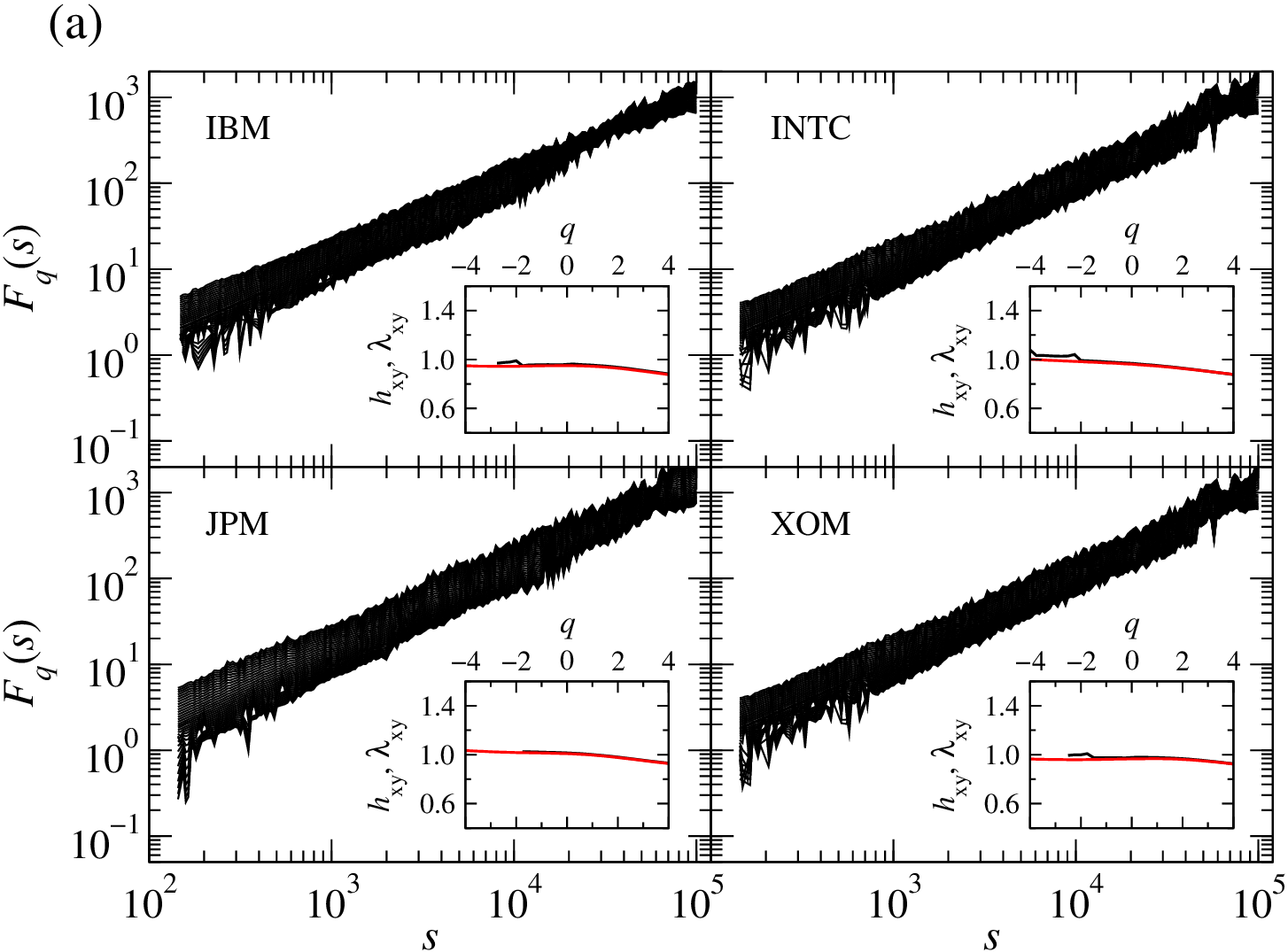}

\vspace{0.8cm}
\includegraphics[scale=0.32]{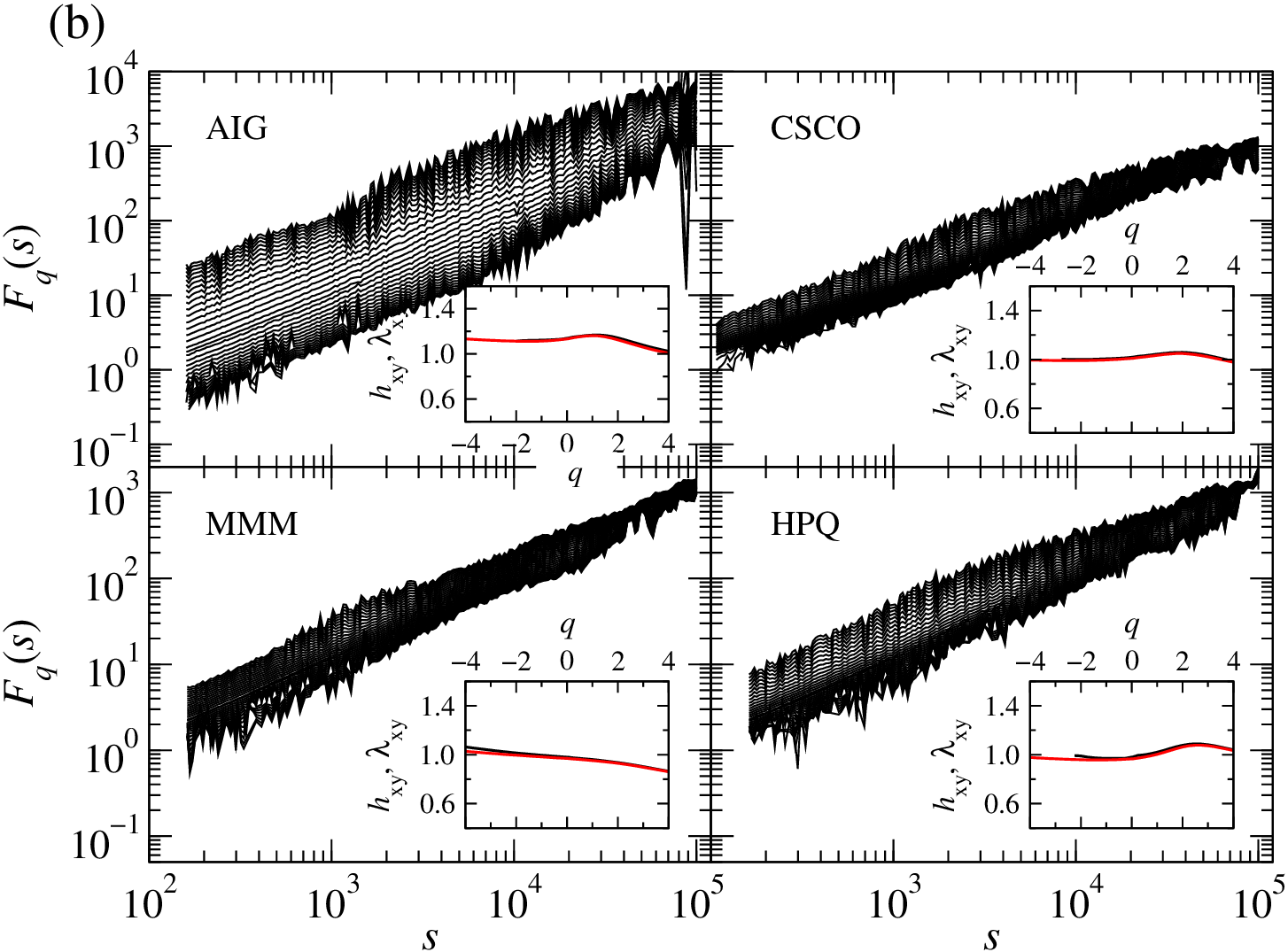}
\caption{The cross-correlation fluctuation function $F_{qXY}(s)$ (main plots) and the scaling exponents: $h_{xy}(q)$ (red/grey, insets) and $\lambda(q)$ (black, insets) calculated for time series of trading activity and volume traded with $\Delta t=1$ min and with $-4 \le q \le 4$ for different stocks: (a) sample stocks with well-defined scaling behaviour of the cross-correlations and (b) sample stocks with less clear power-law cross-correlations. Negative values of $F_{qXY}(s)$, which occur for a few scales $s$ if $q$ is negative, were removed in order to make the plots more readable. In such a case, the respective exponents $\lambda(q)$ are undefined and not shown here.}
\label{fig::mfcca.tv.1min}
\end{figure}

In the remaining part of this Section, we will concentrate on the pair with the best-defined fractality: $T$-$V$. However, even in this case there are strong differences between the results corresponding to different stocks. Sample stocks that exhibit well-defined power-laws in their $T$-$V$ cross-correlations are collected in Fig.~\ref{fig::mfcca.tv.1min}(a). The scaling in each case extends over at least two decades of $s$ and the corresponding exponents $h_{xy}(q)$ and $\lambda(q)$ largely overlap. Sometimes it happens that the results for a few of the smallest (i.e., the most negative) values of $q$ are omitted due to their unstable behaviour for the smallest scales. On the opposite end, there are stocks for which the scaling is somehow distorted. The results of this kind are shown in Fig.~\ref{fig::mfcca.tv.1min}(b). In each case an approximate power-law behaviour may be found in some range of $s$, but the quality of a fit is much poorer than in Fig.~\ref{fig::mfcca.tv.1min}(a). However, this issue does not seem to influence the satisfactory compliance of $h_{xy}(q)$ and $\lambda(q)$.

Typically, the range of the scaling exponent's variability $\Delta \lambda(q)$ in $T$-$V$ is approximately 0.1 and does not exceed 0.2 (approaching this value in few cases only). This makes drawing any decisive conclusion about mono- or multifractality of the data difficult if not impossible, because such values of $\Delta \lambda(q)$ can be obtained either if the cross-correlations are actually monofractal with only spurious multifractality due to the broad probability distribution functions of the data (see~\cite{drozdz2009} for the related discussion) or if there exists actual but weak multifractality. In the case of $\lambda(q)$ being not monotonuous (seen in three panels of Fig.~\ref{fig::mfcca.tv.1min}(b)), the type of fractality is impossible to be defined at all.

\begin{figure}
\includegraphics[scale=0.32]{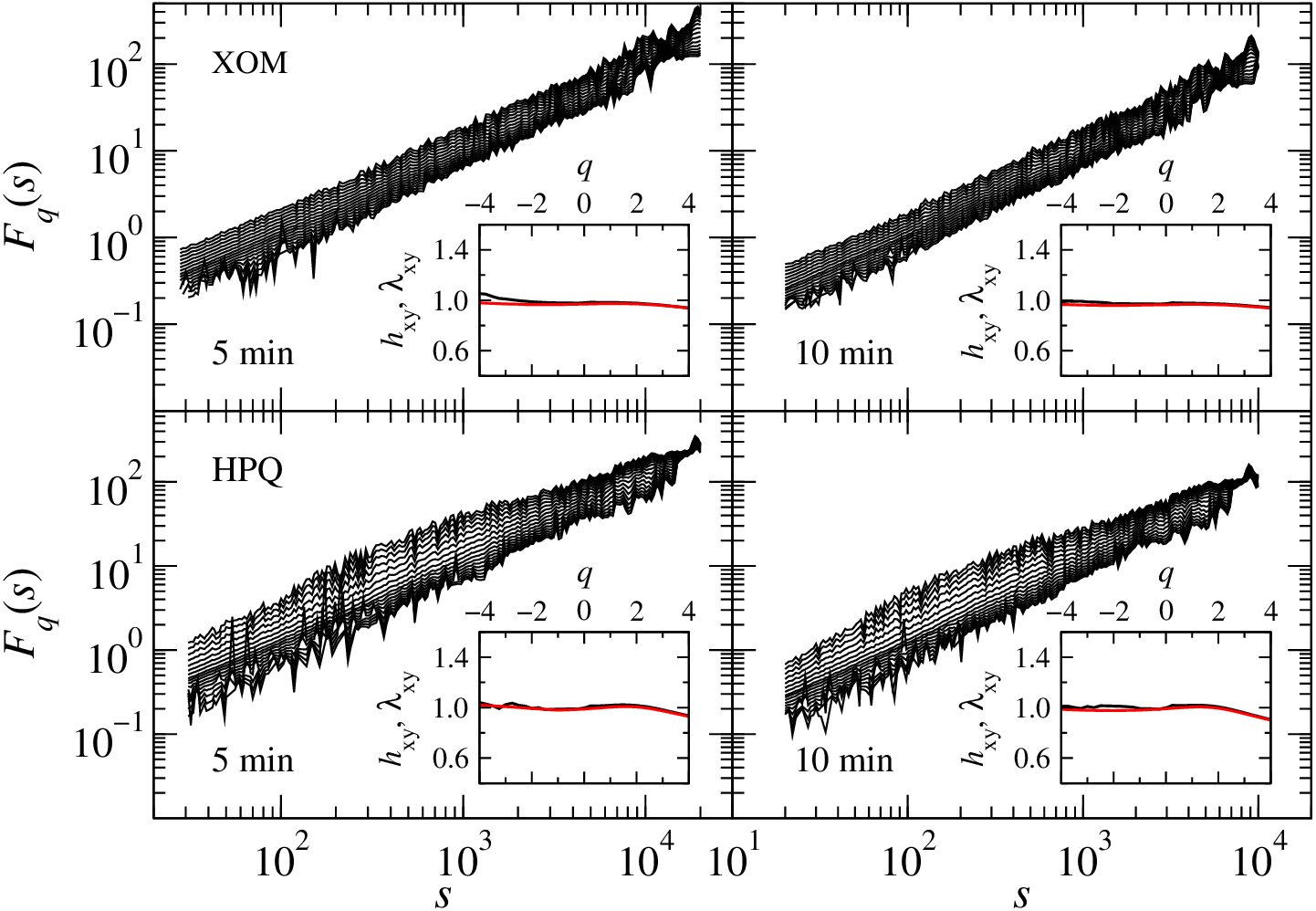}
\caption{The cross-correlation fluctuation functions $F_{qXY}(s)$ (main plots) and the scaling exponents: $h_{xy}(q)$ (red/grey, insets) and $\lambda(q)$ (black, insets) calculated for time series of trading activity and volume with $\Delta t=5$ min and $\Delta t=10$ min and with $-4 \le q \le 4$ for two representative stocks: XOM (well-defined scaling, top) and HPQ (worse scaling, bottom). These results have to be compared with the respective panels of Fig.~\ref{fig::mfcca.tv.1min}. Negative values of $F_{qXY}(s)$ were removed.}
\label{fig::mfcca.tv.scales}
\end{figure}

The above results were obtained for the shortest of the considered temporal scales $\Delta t=1$ min. However, the longer time scales: $\Delta t=5$ min and $\Delta t=10$ min bring results that are qualitatively similar. Fig.~\ref{fig::mfcca.tv.scales} shows the fluctuation functions $F_{qXY}(s)$ calculated for such longer scales in the case of two sample stocks: XOM and HPQ. These results have to be put in the context of the respective panels of Fig.~\ref{fig::mfcca.tv.1min}. In fact, a comparison of the behaviour of $F_{qXY}(s)$ for different $\Delta t$ indicates that there is little dependence of the quality of scaling on a time scale, at least for the considered range of $\Delta t$. Also both the exponents $h_{xy}(q)$ and $\lambda(q)$ closely resemble each other.

\begin{figure}
\includegraphics[scale=0.32]{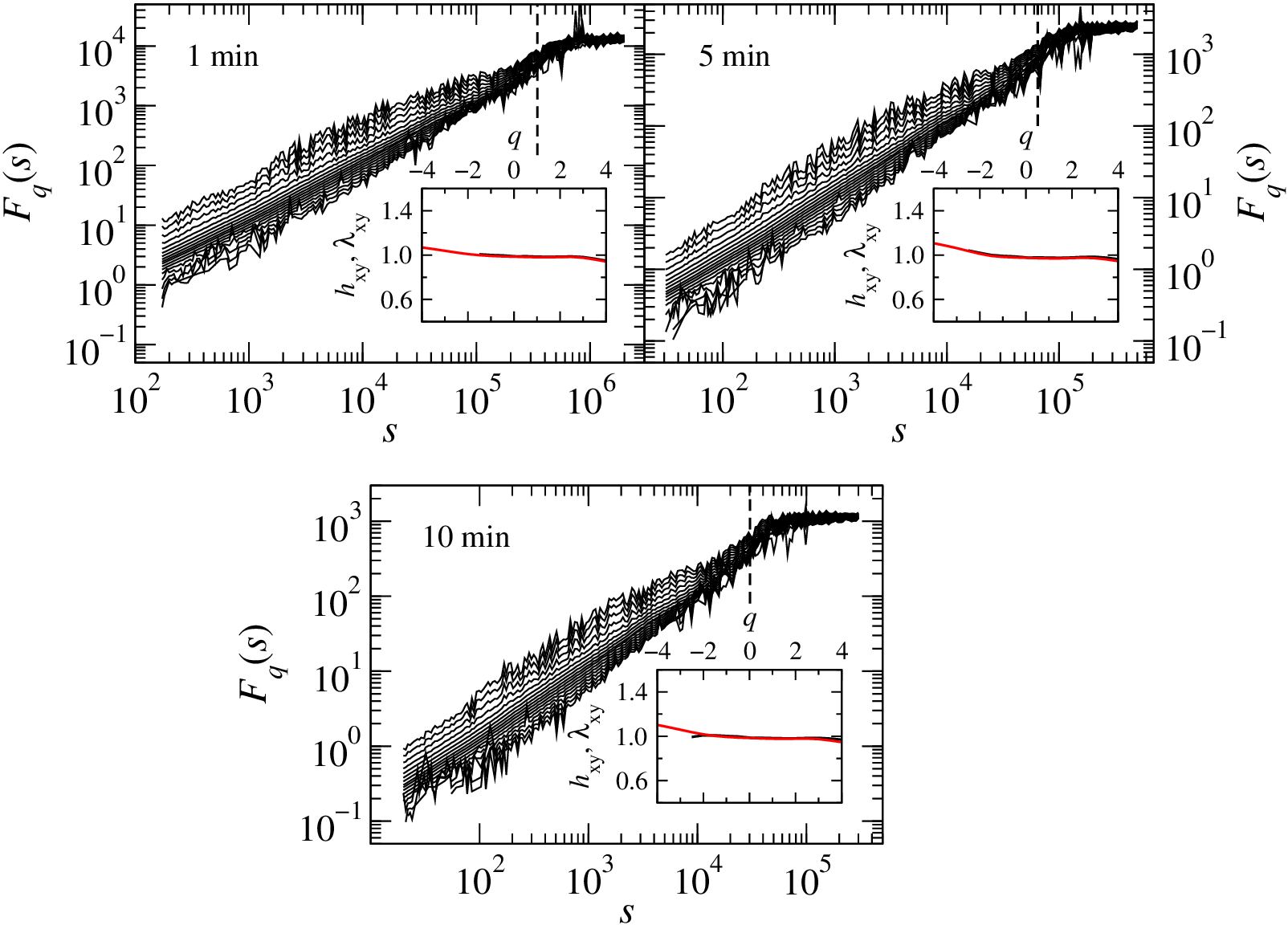}
\caption{The cross-correlation fluctuation functions $F_{qXY}(s)$ (main plots) and the scaling exponents: $h_{xy}(q)$ (red/grey, insets) and $\lambda(q)$ (black, insets) calculated for the joined time series of trading activity and volume traded obtained from $N=31$ stocks with $\Delta t=1$ min (the length: $L^{(N)} \approx 10^7$), 5 min ($L^{(N)} \approx 2 \times 10^6$), and 10 min ($L^{(N)} \approx 10^6$) calculated for $-4 \le q \le 4$. Negative values of $F_{qXY}(s)$ were removed like in Fig.~\ref{fig::mfcca.tv.1min}.}
\label{fig::mfcca.tv.all-stocks}
\end{figure}

Finally, we join the time series representing the same quantity (either $T$ or $V$) and the same time scale $\Delta t$ for all the 31 stocks and in this way we obtain much longer signals that can be used to determine the average behaviour of the cross-correlations between these two quantities. The outcomes, presented in Fig.~\ref{fig::mfcca.tv.all-stocks}, indicate that the quality of the average scaling resembles the plots shown in Figs.~\ref{fig::mfcca.tv.1min}-\ref{fig::mfcca.tv.scales} with a clear improvement if one goes from the shorter to the longer $\Delta t$, a feature that is hardly detectable in the results obtained for the individual stocks. The functions $F_{qXY}(s)$ show also that for $s > L$ their plots tend to flatten and the scaling exponents $\lambda(q)$ decrease to zero. Such behaviour is triggered by the existence of seasonal components~\cite{podobnik2009a,ludescher2011b}, which here are related to the long-term trends in the market evolution that influence majority of the stocks. Such trends cannot be completely removed by the $m$th-degree polynomial $P^{(m)}$ in Eq.~(\ref{eq::detrended.signal}) and they thus survive the detrending procedure and manifest themselves the stronger, the stronger is the inequality $s > L$. As expected, there is no difference between $h_{xy}(q)$ and $\lambda (q)$.

\section{Summary}

We studied high-frequency data from the American stock market comprising 31 stocks listed in DJIA over a 3.5-years-long period between 2008 and 2011. Four quantities were of interest to us: logarithmic returns $R$, volatility $|R|$, trading activity $T$, and volume traded $V$. In the preliminary phase, the time series were detrended in order to remove daily seasonalities. Then, in the proper phase of our study, we confirmed by means of the MFDFA method that the time series representing all these quantities are at least approximately power-law autocorrelated with the Hurst exponent $h(q) > 0.5$ for $q=2$. Next we applied the MFCCA method to pairs of the signals representing the same stock but different quantities and found that the strongest power-law cross-correlations are seen for trading activity and volume, while the weakest ones for returns and any of the other three quantities. The pairs comprising volatility and trading activity or volatility and volume presented the behaviour that situated them somewhere in between. These results were typical from the qualitative perspective, but there were also quantitative differences among the stocks in the quality of the power-law scaling or in the fractal or multifractal properties. We also found that these results were roughly invariant under a change of time unit $\Delta t$ from 1 min to 10 min.

Our observation that the highest power-law cross-correlations occur between trading activity $T$ and volume traded $V$ reveals the existence of strong statistical coupling between these quantities. In particular, it also supports a view that high volumes are produced by a high number of small trades rather than by large individual ones, as one of the assumptions of the model by Gabaix et al. states~\cite{gabaix2003,gabaix2006}.

\section{Acknowledgements}

This work was partially supported by the Centre for Innovation and Transfer of Natural Sciences and Engineering Knowledge (University of Rzesz\'ow).

\end{document}